\documentclass[reprints,secnumarabic,amssymb,notitlepage, nobibnotes, nofootinbib,aps, prd]{revtex4-1}

\usepackage{bm}           
\usepackage{graphicx}
\usepackage{amsmath,amssymb,mathrsfs}  
\usepackage[pdftex,bookmarks,colorlinks,breaklinks]{hyperref}  
\usepackage{bm}
\usepackage{color}
\usepackage{mathrsfs}
\newcommand{\gapp}{\mathrel{\vcenter{\hbox{\tiny\ooalign
{\raise 3.25pt\hbox{$>$}\crcr $\sim$}}}}}

\renewcommand{\mathbf}[1]{{\mbox{\boldmath $\mathrm{#1}$}}}

\newcommand{\be}{\begin{equation}}
\newcommand{\ee}{\end{equation}}
\newcommand{\ba}{\begin{eqnarray}}
\newcommand{\ea}{\end{eqnarray}}

\newcommand{\bac}{\begin{equation}
    \begin{array}{rcl}}
\newcommand{\eac}{\end{array}\end{equation}}

\newcommand{\sfrac}[2]{{\textstyle\frac{#1}{#2}}}

\newcommand{\forget}[1]{\iffalse#1\fi}
\newcommand{\forgetmenot}[1]{\iftrue#1\fi}

\renewcommand{\div}{\hskip0.9pt{\mathsf{div}\hskip2pt}}
\newcommand{\curl}{\hskip0.9pt{\mathsf{curl}\hskip2pt}}

\renewcommand{\S}{_{\hskip-0.8pt\mathrel{\vcenter{\hbox{\tiny\ooalign
{\raise 1.5pt\hbox{\textsf{S}}}}}}}}
\newcommand{\V}{_{\hskip-0.8pt\mathrel{\vcenter{\hbox{\tiny\ooalign
{\raise 1.5pt\hbox{\textsf{V}}}}}}}}
\newcommand{\T}{_{\hskip-0.8pt\mathrel{\vcenter{\hbox{\tiny\ooalign
{\raise 1.5pt\hbox{\textsf{T}}}}}}}}

\begin{document}
\title{The Weyl Curvature Tensor, Cotton-York Tensor and Gravitational Waves: A covariant consideration}
\author{\vspace{2mm} Bob~Osano}
\email{bob.osano@uct.ac.za} \affiliation{Astrophysics, Cosmology and Gravity Centre, Department of Mathematics and Applied Mathematics, University of Cape Town, Rondebosch 7701, Cape Town, South Africa}

\affiliation{Academic Development Programme, Science, Centre for Higher Education Development, University of Cape Town, Rondebosch 7701, Cape Town, South Africa}

\date{\small{\today}}
\begin{abstract}
\section{Abstract}
1+3 covariant approach to cosmological perturbation theory often employs the electric part ($E_{ab}$), the magnetic part ($H_{ab}$) of the Weyl tensor or the shear tensor ($\sigma_{ab}$) in a phenomenological description of gravitational waves. The Cotton-York tensor is rarely mentioned in connection with gravitational waves in this approach. This tensor acts as a source for the magnetic part of the Weyl tensor which should not be neglected in studies of gravitational waves in the 1+3 formalism. The tensor is only mentioned in connection with studies of 'silent model' but even there the connection with gravitational waves is not exhaustively explored. In this study, we demonstrate that the Cotton-York tensor encodes contributions from both electric, magnetic part of the Weyl tensor and in directly from the shear tensor. In our opinion, this makes the Cotton-York tensor arguably the natural choice for linear gravitational waves in the 1+3 covariant formalism. The tensor is cumbersome to work with but that should negate its usefulness. It is conceivable that the tensor would equally be useful in the metric approach, although we have not demonstrated this in the current study. We contend that the use of only one of the Weyl tensor or the shear tensor, although phenomenologically correct, leads to loss of information. Such information is vital particularly when examining the contribution of gravitational waves to the anisotropy of an almost -Friedmann-Lamitre-Robertson-Walker (FLRW) universe. The recourse to this loss is the use Cotton-York tensor. 

\end{abstract}
\maketitle
An important deduction from Einstein's theory of general relativity is that no speeds faster than the speed of light are admissible. This limitation suggests that any changes in the gravitational field must, utmost, propagate as gravitational waves. Such waves transport energy, in the form of radiation, associated with the changes in the gravitational field. The nonlinear nature of general relativity makes the task of characterizing of gravitational waves non trivial. Indeed, only in certain approximations can one clearly define gravitational radiation. There are three approximations in which it is possible to make this definition: the linearized theory ( weak field approximation where small perturbation about a nearly flat model are considered. we note that the question of the generation of the wave cannot be answered in this approximation), post-Newtonian theory (which is good for describing sources where waves arise as higher order correction and encodes GR) and perturbation theory \cite{Cal}. This paper examines gravitational radiation in the context of cosmological perturbations wherein two main approaches feature prominently. These are: (1) the metric approach and (2) the covariant approach. Studies of gravitational waves in these two approaches rely on tensors that are projected, symmetric and trace-free. The focus of this article is the Weyl tensor, and the role it plays in describing gravitational waves in the covariant approach. Related to the Weyl tensor is the 3-Cotton-York (hereafter $C_{ab}$) tensor. We investigate the $C_{ab}$ tensor in order to determine its potential for gravitational waves description.

A local definition of gravitational-radiation in the language of geometry of space-time was first given in \cite{ADM} with gravitational waves understood as the mechanism via which radiation energy is transported\footnote{This was pointed out to the author by S. Deser, he of the 'ADM'}. In particular, a link between the conformal curvature tensor of 3-surfaces, and the existence of gravitational waves was drawn. It was later argued that conformal tensor could be taken as a measure of radiation amplitude \cite{York}. The authors of \cite{Berger} argued that the vanishing of Cotton-York tensor should be taken as indicative of a non-radiative nature of the space-time, much in line with vanishing of the magnetic part of the Weyl tensor leading to the 'silent' models \cite{Mat}.

Although we know that the definition of gravitational radiation is dependent on the space-time slicing, it is tempting to ask if this is generic to models that admit similar slicing. In particular, is the non-vanishing of Cotton-York tensor an indicator of the existence of gravitational waves in a model? Other analyses of the Cotton-York tensor in relation to gravitational radiation have been done in \cite{Lukash, Wainright, Berger, ThesisHenk}. The need to investigate whether or not the algebraic structure of the Cotton-York tensor is related to the nature of the gravitational waves was pointed out in \cite{Wainright}. This article is a part of these investigations.  We use the 1+3 covariant approach to general relativity and cosmology, which date back to the 50s \cite{heck, Ray}. 

\section{The 1+3 perturbations formalism and general equations}
In the covariant-gauge invariant 1+3 formalism, the fundamental variable is the 4-velocity; $u^{a}$ taken to represent the motion of the fundamental observer. We take this velocity to be defined by the cosmic microwave radiation background, but observe that there are other ways of defining it. Since we are interested in a hyper-surface against which all dynamics will be considered, we define a projection tensor $h_{ab}=g_{ab}+u_a u_b$, which under suitable condition will projected onto a hyper-surface that we are interested in. We also define a spatial derivative $D_{a}= h^{c}_{a}\nabla_{c}$, where $\nabla_{c}$ is the derivative used in the metric approach. These operators together with the 4-velocity and the energy-momentum tensor allow for the definition of variables that describe the dynamics in the background given by the hypersurface. In particular, the rate of expansion: $\Theta=D_{a}u^{a}$ and the energy density $\mu=T_{ab}u^{a}u^{b}$ and isotropic pressure $p=\frac{1}{3}T_{ab}h^{ab}.$ A detailed account of the 1+3 formalism can be found in \cite{EMM, EB89}. 

The concise perturbation theory in the 1+3 formalism which was first presented in \cite{EB89}. This approach is, figuratively speaking, a {\it top-down} approach. It begins with the development of the bigger picture given in terms of non linear systems of propagations and constraint equations representing the fully perturbed system. Such equations are then linearized about a background taken as representing the non perturbed model. This approach differs from the standard gauge-invariant metric perturbation theory that can be thought of as a {\it bottom-up} approach.  In the latter, one begins with variables representing a background of choice, for example the metric, that is then perturbed to the desired order before being propagated. In both approaches, the notion of perturbed model, unperturbed model and the transformation relationships between variables in the two model is critical. This has been discussed in great detail in a number of articles, see the following for example \cite{ Bardeen, Mukhanov, Kodama, EB89}.

In all approaches to cosmological perturbations, it is easy to demonstrate that the various types of perturbations; namely scalar, vector and tensor, decouple when either linearization is done ( only terms of magnitude up to first order are considered) as is the case in the 1+3 formalisms, or equations are perturbed to first order as is the case in the metric approach. This natural decoupling does not occur in higher order perturbations as has been shown in a number of articles, for example \cite{Matt, Nak, Bob2007, Bob2017, CB1}.  In the present study, we are interested in linear tensor perturbations capable of describing gravitational waves. The linear order tensor perturbations about a FRLW background, for example, are easy to define in the 1+3 formalism. These are the shear tensor $\sigma_{ab}=D_{\langle a}u_{b\rangle}$; where the angle brackets indicate projected indices, the electric part of the Weyl tensor $E_{ab} (\equiv C_{cdgh}{C^{gh}}_{ef}{h^{c}}_{a}u^{d}{h^{e}}_{b}u^{f}$ where $C_{cdef}$ is the Weyl tensor) and the magnetic part of the Weyl tensor $H_{ab}(\equiv \frac{1}{2}\varepsilon_{cdef}{h^{c}}_{a}u^{d}{h^{e}}_{b}u^{f})$, where $\varepsilon_{cdef}$ is a permutation tensor. These tensors represent an effect on the geometry. We use units with $c=1$ (speed of light) and $8\pi G=1$ throughout. The geometry-matter interaction is given by the Einstein gravitational field equations: \begin{eqnarray}R_{bc}-\frac{1}{2}Rg_{bc}=T_{bc},\end{eqnarray} which is given for vanishing cosmological constant, i.e. $ \Lambda=0$. $R_{bc}$ is the Ricci tensor, $R$ is the Ricci scalar and $g_{bc}$ is the metric tensor. From {\it Ricci identities} associated with the 4-velocity vector field $u^{a}$ (namely: $2\nabla_{[a}\nabla_{b]}u^{c}={{R_{ab}}^{c}}_{d}u^{d}$), one can derive the {Raychaudhuri equation,} the vorticity propagation equation, the shear propagation equations, the shear divergence constraint, vorticity divergence identity constraint and the magnetic Weyl tensor constraint. Of these, only the Raychaudhuri equation (\ref{Ray}) tells us something about the background dynamics, while the others tell us something about perturbation related dynamics.

\subsection{Background dynamics} The full nonlinear Raychaudhuri equations is
 \begin{eqnarray}\label{Ray}\dot{\Theta}=-\frac{1}{3}\Theta^2-\frac{1}{2}(\mu+3p)-2\sigma^2+2\omega^2+D_{a}\dot{u}^{a}+\dot{u}^2,\end{eqnarray} where $\sigma^{2}\equiv\sigma_{ab}\sigma^{ab} $, $\omega^{2}\equiv\omega_{ab}\omega^{ab}$ and $\dot{u}^2\equiv\dot{u}_{a}\dot{u}^{a}$. Since the definition of orthogonal projection onto hypersurfaces demand that $\omega_{ab}=0$, it follows that $\omega^{2}=\sigma_{ab}\sigma^{ab}=0.$ We note that this equation is composed of parts that represent background dynamics ( Zeroth order), parts that represent linear order dynamics (First-order) and parts that are clearly nonlinear (Second-order, given that  they are products of first-order quantities). Examples of the nonlinear terms are $\sigma^2, \omega^2$ and $\dot{u}^2$ (the reader is referred to \cite{Bob2017} for the basis for the classification of such variables into zeroth, first- or second-order. It is instructive to state that besides the notion of 'smallness' with respect to amplitude $\mathcal{O}(\epsilon)$ being at the foundation of this classification, so is the notion existence or non-existence of a quantity at one particular order). To be precise, two quantities that vanish in the background will each be of order $\mathcal{O}(\epsilon)$. The product of such two quantities will have order $\mathcal{O}(\epsilon^2)$. In comparison, a quantity that vanishes in both the background and first-order, but not second-order, will have order $\mathcal{O}(\epsilon^2)$. These considerations inform perturbation scheme used in the 1+3 formalism. In particular, the scheme allows for the analysis of zeroth-order, first-order and second-order dynamics ( the reader is referred \cite{EB89, CC1, Bob2017} for issues that are related to 1+3 perturbations scheme and how some are resolved). In this article, we are only concerned with the zeroth-order ( unperturbed background space-time) and the first-order (perturbations to linear order). Background dynamics require $\sigma^2=D_{a}\dot{u}^{a}=\dot{u}^2 =0$ leaving \begin{eqnarray}\label{thetadot}\dot{\Theta}=-\frac{1}{3}\Theta^2-\frac{1}{2}(\mu+3p).\end{eqnarray} 

The twice-contracted Bianchi identities ensures the total energy-momentum is conserved, while the energy conservation equation leads to the following propagation equation in the background; 
 \begin{eqnarray}\label{mudot}\dot{\mu}=-(\mu+p)\Theta.\end{eqnarray} We often think of $\sigma_{ab}, E_{ab}, H_{ab}$ in totality as tensors but, as we shall see, they may be split into pure scalar, pure vector and pure tensor parts with each obeying their respective transformation property. Indeed the pure tensor part transforms as a tensor. For now, it is prudent to refer to such quantities as pseudo-tensor or {\it scalar-vector-tensor}, in the context of general relativity and to use the notation (SVT) a reminder that they are not necessarily pure tensors. We will return to this splitting later and particularly to the discussion of tensorial part in relation to gravitational waves. At this stage, we can present a set of linearized propagation equations for these quantities. These are the tensorial differential equations :

\ba
\label{S}\dot{\sigma}_{ab}+\frac{2}{3}\Theta\sigma_{ab}+E_{ab}&=& D_{\langle a}\dot{u}_{b\rangle}\\
\label{E}\dot{E}_{ab}+\Theta E_{ab}-\curl H_{ab}-\frac{1}{2}(\mu+p)\sigma_{ab}&=&0\\
\label{H}\dot{H}_{ab}+\Theta H_{ab}+\curl E_{ab}&=&0
\ea
where $\curl \sigma_{ab}\equiv\varepsilon_{ef\langle a}D^{e}{\sigma_{b\rangle}}^{c}$ (the permutation tensor $\varepsilon_{abc}=u^{d}\eta_{abcd}$ is again a volume element \cite{EHvE98}. The complete nonlinear equation for $\dot{\sigma}_{ab}$ is derived from the {Ricci identities} and then linearized to get equation (\ref{S}). Similarly, the full nonlinear equations for $\dot{E}_{ab}$ and $\dot{H}_{ab}$ are derived from the Bianchi identities $\nabla_{[a}R_{bc]de}=0$ then linearized to get (\ref{E}) and (\ref{H}). In particular, one can split the Riemann tensor $R_{abcd}$ into the Ricci tensor $R_{ab}$ and the Weyl curvature tensor $C_{abcd}$. The 1 + 3 splitting of these quantities, and EFE, together with the once-contracted Bianchi identities generate nonlinear equations for $\dot{E}_{ab}$ and $\dot{H}_{ab}$. A detailed discussion of this is given in \cite{EHvE98}. The propagation equations (\ref{S}, \ref{E}, \ref{H}) are accompanied by the following linearized constraint equations:
\begin{eqnarray} \label{CS}D^{b}\sigma_{ab}&=& \frac{2}{3}D_{a}\Theta\\
\label{CH}\curl\sigma_{ab}&=& H_{ab}\\
\label{CE}D^{b}E_{ab}&=&\frac{1}{3}D_{a}\mu\\
\label{DP}\nabla_{a}p&=&(\mu+p)\dot{u}_{a}.\end{eqnarray} It is clear, from the right hand side of equations (\ref{CS}, \ref{CE} and \ref{DP}), that the set of propagation and constraints equations is made of terms that have the properties of the previously mentioned {\it scalar-vector-tensor} (SVT) and others that have {\it vector-tensor} (VT) properties. In order to gain a better understand of what this means, we need a brief review of covariant spitting in the 1+3 formalism. To this end, it is clear that zeroth-order (background) dynamics is determined by $\mu, p$ and $\Theta $   ( which are non-zero for the case where back-reaction is neglected). This means that in the background; $D_{a}\mu, D_{a}\Theta, \textrm{E}_{ab}, \textrm{H}_{ab},$ and $\sigma_{ab}$ all vanish as they are all first-order ( i.e. $\mathcal{O}(\epsilon)$. An example of this background is the FLRW space-time. We now discuss the splitting of the first-order variables.\par

\subsection{Covariant Splitting} Consider a background that is spatially homogeneous and isotropic, against which all quantities are defined. It follows that each first-order vector quantity, denote by $V_{a}$, may be split uniquely into the curl-free (the scalar) and the divergence-free (the vector) parts. i.e. $V_{a}=V^{S}_{a}+V^{V}_{a}$ where $\curl V^{S}_{a}=0$ and $\div V^{V} =0$. Any tensor $W_{ab}$ may also be invariantly split into scalar, vector and tensor parts i.e.$W_{ab} =W^{S}_{ab} +W^{V}_{ab} +W^{T}_{ab}$ where $\curl W^{S}_{ab} =0$, $\div\div W^{V} =0$, and $\div W^{T}_{a} =0$ respectively. Implementing such splits in equations ({\ref{CS}-\ref{CH}) allow the comparison of terms on one side of the equal sign to those of the opposite side. We reiterate that rotation is ignored given the requirement for the existence of a hyper-surface. This implies that vectors part effectively set to vanish. Taken together, these splits enable us to understand why $H_{ab}$ is the natural suitable quantity for the covariant characterization of gravitational waves in the 1+3 formalism this far. But the splitting  also give a hint of why contributions from other tensors should not be neglected, as is presently the case. Lets examine this by looking at a number of equations.
\subsubsection{The Weyl tensor and gravitational waves} 
We note that equation (\ref{CS}) gives \ba \label{D1}D^{b} (\sigma^{S}_{ab}+\sigma^{V}_{ab}+\sigma^{T}_{ab})=\frac{2}{3} (D_{a}\Theta)^{S}+\frac{2}{3} (D_{a}\Theta)^{V},\ea which means that $D^{b} \sigma^{T}_{ab} =0$, as we would expect. But in order to isolate the tensor part, one should take the curl of equation (\ref{D1}) and use the relevant commutation\cite{RM1,HVE1} relation, with the scale factor correctly accounted for, to get \[\curl (D^{b} \sigma_{ab})=2D^{b}(\curl\sigma_{ab})=2D^{b}\curl(\sigma^{T}_{ab})=0,\]which is effectively the divergence of equation (\ref{CS}). We see that $\sigma^{T}_{ab}$ sources the divergence-free $H_{ab}$. It is easy to show by taking the time derivative of equation (\ref{S}), using equation (\ref{E}) and a linearized commutation 
\ba\label{com1}(\curl T_{ab})^{.}=\curl\dot{T}_{ab}-\frac{1}{3}\Theta\curl{T}_{ab}, \ea that $H_{ab}$ forms a closed wave equation: \ba\ddot{H}_{ab}-D^{2}H_{ab}+\frac{7}{3}\Theta\dot{H}_{ab}+(\frac{2}{3}\Theta^2 - 2p)H_{ab}=0.\ea As shown in \cite{Challinor}, one can express this in mode expansion by using the following parity tensor harmonics, $H_{ab}=l^{-2}\Sigma_{\kappa}(\kappa)^2(H_{\kappa}Q^{(k)}_{ab}+ \bar{H}_{\kappa}\bar{Q}^{(k)}_{ab})$, where $Q_{ab}, \bar{Q}_{ab}$ are the dual tensor harmonics for the different parities and $l$ the scale-factor. The expansion for the electric parity yields \begin{eqnarray}\label{HL}\ddot{H}_{\kappa}+\frac{5\dot{l}}{l}H_{\kappa}+\left(\frac{\kappa^{2}}{a^{2}}-2(\frac{\dot{a}}{a})^{2}+2\frac{\ddot{a}}{a}-2p\right)H_{\kappa}=0.
\end{eqnarray} The magnetic parity will have a similar structure. Since we are not intending to solving this equation but rather to compare it to the gravitational wave equation in other formalism, it suffices to say that the two scale factor terms in the last bracket can be replaced using Friedmann equations.
 This is comparable to the tensor perturbations, $H^{(2)}_{T}$, in the Bardeen's formalism \cite{Bardeen} namely
\begin{eqnarray}\label{HB}\ddot{H}^{(2)}_{T}+\frac{2\dot{l}}{l}H_{T}+ (\kappa^{2}+2K)H_{T}=0,
\end{eqnarray} where $l$ is the scale factor, $K$ is the spatial curvature of the FRLW model and $\kappa$ is the wave number (which has a physical significance when $K$=0). Please note that $H_{\kappa}$ and $H_{T}$ are not the same tensor. This aside, the analogous nature of the two tensor equations makes the magnetic part of the Weyl tensor a natural choice for characterizing gravitational waves in the 1+3 formalism \cite{Lesame1, Lesame2}. But one would argue that $\sigma_{ab}$ has these very characteristics, so why not use $sigma_{ab}$? The answer to this question lies in equation (\ref{D1}) where it is clear that divergence of the full shear tensor is sourced by the gradient of expansion which has a scalar contribution. This, however, does mean that $\sigma^{T}_{ab}$ cannot characterize gravitational waves. If one can extract the pure tensor part (which can be done \cite{CC1}), one can indeed characterize gravitational waves using the $\sigma^{T}_{ab}$. What about $E_{ab}$? It is often said that for gravitational waves to exist we require that $D^{b}H_{ab}=0$ and $D^{b}E_{ab}=0.$ One would then expect $E_{ab}$ to play a role similar to that of $H_{ab}$ in the description of gravitational waves. This being said, $E_{ab}$ has the following characteristics;\begin{enumerate}
\item The divergence is sourced by density perturbations as in equation (\ref{CE}), and 
\item It does not form a closed second order wave equation i.e \[\ddot{E}_{ab}-D^{2}E_{ab}+\frac{7}{3}\Theta\dot{E}_{ab}+\frac{2}{3}\Theta^2 E_{ab}=\frac{1}{6}\mu\Theta\sigma_{ab},\]
\end{enumerate} unless one considers an empty ($\mu=0$) universe as was done in \cite{Hawking}. $E_{ab}$, which could be expanded using tensor harmonics \cite{Challinor}, is linked to $\sigma_{ab}$ which sources $H_{ab}$ that characterizes gravitational waves, does it not following that we loose information by not considering the contribution of $E_{ab}$ to the gravitational waves? 

We note, from equation (\ref{CE}), that \[D^{b}E^{S}_{ab}+D^{b}E^{V}_{ab}+D^{b}E^{T}_{ab}=\frac{1}{3} (D_{a}\mu)^{S}+\frac{1}{3} (D_{a}\mu)^{V},\] which implies $D^{b}E^{T}_{ab}=0$ as is expected. Also note that taking the curl of this equation yields $\curl D^{b} E_{ab}=0=2 D^{b}\curl E_{ab}.$ Although one cannot write down a closed wave equation for $E_{ab}$, the effective order of the equation is two since $E_{ab}$ can be determined from the shear equation (\ref{S}), and the shear is determined by a wave equation \cite{Challinor}. So why not incorporate the effect of $\curl E_{ab}$? Rather than considering $H_{ab}$ and $\curl E_{ab}$ separately, it is sufficient to find a tensor that captures the contributions from both tensors, hence the $^{3}$Cotton-York tensor  which we denote using $C_{ab}$.

\subsubsection{The Cotton-York tensor and gravitational waves} Let $^{3}\textrm{S}_{ab}$ be the trace-free part of the 3-Ricci curvature tensor that is given by Gauss - Codacci equations for the Ricci curvature of the spatial surfaces. one can show that $S_{ab}$ is relates to the Ricci scalar and tensor in the following manner \cite{GE1}\ba ^{3}\label{eqn1}\textrm{S}_{ab}={^{3}\textrm{R}}_{{ab}}-\sfrac{1}{3}{^{3}\textrm{R}h_{ab}}~,\ea
where $^{3}\textrm{R}_{ab} $ and $^{3}\textrm{R}$ are the Ricci tensor and Ricci scalar respectively. It is easy to show that \ba ^{3}\textrm{S}_{ab}&=&-\frac{1}{3}\Theta\sigma_{ab}+E_{ab}\ea to first order. It is also obvious that in taking the curl of this equation and using the relevant commutation relation, one obtains the link between $H_{ab}$, $\curl E_{ab}$, and $\curl ^{3}S_{ab}$, a link that we will exploit in this analysis. It is known that $\curl ^{3} S_{ab}$ is proportional to the $C_{ab}$ i.e. $\label{eqn0}{^{3}\textrm{C}}_{ab}=h^{1/3}\curl \textrm{S}_{ab},$ where $h$ is the modulus of the projection tensor $h_{ab}$ ( i.e ${h^{a}}_{b}={\delta^{a}}_{b}+u^{a}u_{b}$, such that $h_{ab}u^{a}=0$). We can express $C_{ab}$ in terms of the other variables whose propagation and constraint equations are known. This allows for ease of calculations and interpretation of results. The full non-linear $C_{ab}$ tensor in terms of other variables takes the form;
\ba\label{C_{ab}} \textrm{C}_{ab}&=&-h^{1/3}[\dot{{H}}_{ab}+\frac{4}{3}\Theta{H}_{ab}+\mathcal{O}(e^{2})]\ea where $\mathcal {O}(e^2)$ represents a sum of terms that include $3\sigma_{c\langle a}{{H}^{c}}_{b\rangle}-\varepsilon_{cd\langle a}\left(D^{c}(\sigma^{de}\sigma_{b\rangle e})+\frac{1}{3}(D^{c}\Theta){\sigma_{b\rangle}}^{d}\right)$. These terms are dropped in the linearization scheme. In practice, the full nonlinear equation is derived and then linearized. It is clear that this tensor has first-order and second-order parts that are given by the quadratic products of first-order variables. We now show that $C_{ab}$ tensor is divergence-free. \\\\
The divergence of $C_{ab}$ is given by $(D^{b}\textrm{C}_{ab} )=h^{1/3}D^{b}\curl(\textrm{S}_{ab})$, where $D^{b}h=0$ by definition. In order to expand the left hand side of this divergence formula, we make use of the commutation relation \ba \label{eqn2}D^{b}\curl{^{3}\textrm{S}_{ab}} =\sfrac{1}{2}\curl D^{b}({^{3}\textrm{S}_{ab}}),\ea where only first-order terms are kept while higher order terms are dropped. The relevant set of commutation relations are given in \cite{RM1, HVE1,ThesisHenk}. From equation (\ref{eqn1}), it follows that \ba\label{eqn3}D^{b}( {^{3}}\textrm{S}_{ab})=D^{b}({^{3}}\textrm{R}_{ab})-\sfrac{1}{3} D_{a} {^{3}}\textrm{R}.\ea It also follows from the internal consistency of the constraint and the evolution equations, subject to the contracted Bianchi identities of 3-surfaces, that $D^{b}({^{3}}\textrm{R}_{ab})=\sfrac{1}{2}  D_{a}{^{3}}\textrm{R}$  (which is equivalent to the part of curvature that is not locally determined by matter\cite{Hawking}). This implies that $D^{b}( {^{3}}\textrm{S}_{ab})=\sfrac{1}{6}  D_{a}{^{3}}\textrm{R}$. We have already indicated that gradients of scalars are first-order in approximation. In fact it easy to demonstrate that for the hyper-surface which is defined by a vanishing vorticity, a 3-Ricci scalar exists such that $^{3}R=-\frac{2}{3}\Theta^2+2\mu+\mathcal{O}(\epsilon^2)$ when the cosmological constant $\Lambda=0$, and where $\mathcal{O}(\epsilon^2)=\sigma_{ab}\sigma^{ab}.$ Note that $\omega_{ab}\omega^{ab}=0$ since vorticity vanishes. This means that 
\ba\label{eqn4}D_{a}({^{3}\textrm{R}})=-\frac{4}{3}\Theta D_{a}\Theta+2D_{a}\mu,\ea up to first-order approximation. Since vector components are being ignored, the curl of the right hand side vanishes, as only scalars are present. This can be seen from taking the curl of equation (\ref{CH}), using a commutation relations and equation (\ref{CS}) where \be D^{b}\textrm{H}_{ab}=D^{b}\curl \sigma_{ab}=\sfrac{1}{2}\curl D^{b}\sigma_{ab}=\sfrac{1}{3}\curl D_{a}\Theta.\ee It is then clear from the example of the magnetic part of the Weyl tensor $H_{ab}$ that $D^{a}H_{ab}=0$ if, and only if, $\curl D_{a}\Theta =0.$ Similarly, $\curl D_{a}\mu=0\Rightarrow D^{b}\curl\textrm{E}_{ab}=0.$ The two conditions: namely $\curl D_{a}\Theta =0$ and $\curl D_{a}\mu=0$, are the very requirements that make the curl of the right hand side of equation (\ref{eqn4}) to vanish and hence for $D^{b}\textrm{C}_{ab}=0$. We conclude that the $\textrm{C}_{ab}$ satisfies the transverse condition for linear-order perturbations when vector modes of perturbations are neglected. Taking the time derivative of equation ({\ref{C_{ab}}), using equation (\ref{H}), equation (\ref{thetadot}) the linear commutation relation given in equation (\ref{com1}) and linearizing the resulting equation leads to the propagation equation for $C_{ab}$ which has the form \ba\dot{\textrm{C}}_{ab}+\Theta \textrm{C}_{ab}-\curl\curl H_{ab}+(\frac{1}{3}\mu -\frac{1}{9}\Theta^2)H_{ab}=0.\ea Taking a second time derivative, using the commutation relation given in equation (\ref{com1}), and the linear $'\curl-\curl'$ identity: 
\ba
\curl\curl C_{ab}&=&-D^{2}C_{ab}+\frac{3}{2}D_{\langle a}D^{c} C_{b\rangle c}+(\mu-\frac{1}{3}\Theta^2)C_{ab},\nonumber\\
\ea gives the tensor equation 
\ba\label{C2}\ddot{\textrm{C}}_{ab}+3\Theta\dot{\textrm{C}}_{ab}-D^{2}\textrm{C}_{ab}+(\sfrac{13}{9}\Theta^2+\sfrac{1}{6}\mu-\sfrac{3}{2}p)\textrm{C}_{ab}=0,
\ea which is the desired closed wave equation for $C_{ab}$. Here too we can defined and use harmonics expansion. In particular, let \ba C_{ab}=l^{-2}\sum_{k}\left(C_{\kappa}Q^{(\kappa)}_{ab}+\bar{C}_{\kappa}\bar{Q}^{(\kappa)}_{ab}\right).\ea Applying this to equation (\ref{C2}) and isolating the electric parity gives
\ba
\label{this}\ddot{C}_{\kappa}+5\sfrac{\dot{l}}{l}\dot{C}_{\kappa}+\left[\left(\sfrac{\dot{l}}{l}\right)^{2}-2\sfrac{\ddot{l}}{l}+\sfrac{1}{6}\mu-\sfrac{3}{2}p+\sfrac{\kappa^2}{l^2}\right]C_{\kappa}=0.
\ea where one can use Friedmann equations to simplify the square bracket. The other parity has a similar equation. This is the Cotton-York tensor wave equation which compares to the magnetic part of the Weyl ($H_{\kappa}$ in equation (\ref {HL}) and used in \cite{Dunsby}) or even the electric part \cite{Challinor} as used in the 1+3 formalism or in \cite{Hawking}. It also compares to the metric tensor perturbation ($H_{\kappa}$ in equation (\ref{HB})) in the Bardeen formalism. The strength of the Cotton-York tensor lies in that  fact that it encodes contributions from both the magnetic ($H_{ab}$) and the electric ($curl E_{ab}$) parts of the Weyl tensor. This means that the information that would otherwise have been lost by choosing only one of the Weyl tensors is preserved by using the Cotton-York tensor. The natural area to apply this is the CMB anisotropy and polarization as was done for $E_{ab}$ and $\sigma_{ab}$ in \cite{Challinor} and \cite{Challinor2} something that we will pursued elsewhere.
In the current context, it is worth noting that the $H_{ab}=0$ is a defining requirement for dust-filled 'silent' models. These are models that, by definition, lack gravitational waves and sound waves. From equation (\ref{C_{ab}}), it is clear that, up to linear order, $H_{ab}=0$ implies that $C_{ab}=0$ and hence the 'silent' condition still holds. The relevance of $C_{ab}$ tensor to gravitational waves has also been considered in the case of 'silent' inhomogeneous models \cite{ThesisHenk}.
\section{\it Conclusion}
We have demonstrated that the 3-Cotton-York tensor, $C_{ab}$, is divergence-free against an almost - FLRW model when vorticity and vectors are ignored. We have also found that $C_{ab}$ gives a closed wave equation. $C_{ab}$, like $E_{ab}$ and $H_{ab}$, is not locally determined by matter field. These properties suggest that $C_{ab}$ is equally suitable for describing gravitational waves in the 1+3 formalism. It is clear from literature that $\sigma_{ab}$, $E_{ab}$ and $H_{ab}$ have all been used to describe gravitational waves, which suggests that there exists no unique tensor for the description. It is our contention that it should be the preferred variable given that it encodes contributions from both the magnetic and electric parts of the Weyl tensor. Since this is a phenomenological description of gravitational waves, we argue that the hidden third temporal derivative of the magnetic part of the Weyl tensor in $C_{ab}$ does not invalidate its utility. $C_{ab}$ should be the more suitable tool when examining the contribution of gravitational waves to the anisotropy in an almost -FLRW model. Although an almost -FLRW model has been considered in this article, the usefulness of $C_{ab}$ is not restricted to this models as will be shown elsewhere\cite{Bian}.\section{{Acknowledgments}} The author wishes to thank the reviewers for the very helpful comments and pointers. This project was funded by NGP program of the University of Cape Town

\end{document}